\documentclass{elsarticle}
\usepackage{graphicx}
\usepackage{newtxtext}
\usepackage{newtxmath}
\usepackage{natbib}
\usepackage{hyperref}
\usepackage{lineno,hyperref}
\usepackage{epstopdf, epsfig}
\modulolinenumbers[5]
\hypersetup{
    colorlinks = true,
    urlcolor   = blue,
    citecolor  = black,
}

\newcommand{\RomanNumeralCaps}[1]
\linenumbers

\begin{document}
\begin{frontmatter}

% {\MakeUppercase{\romannumeral #1}}

\title{Understanding the effects of rotation on the wake of a wind turbine at high Reynolds number}

% \author{Alexander Piqu\'e\aff{1}
%   \corresp{\email{apique@princeton.edu}},
%   M. A. Miller\aff{2}
%  \and M. Hultmark\aff{1}}

% \affiliation{\aff{1}Mechanical and Aerospace Engineering Department, Princeton University, NJ, USA
% \aff{2}Aerospace Engineering Department, Pennsylvania State University, PA, USA}

\author{Alexander Piqu\'e\corref{mycorrespondingauthor}}
\address{Pennsylvania State University, State College, PA 16801, USA}
\cortext[mycorrespondingauthor]{Corresponding author}
\ead{axp5446@psu.edu}

\author{Mark A. Miller}
\address{Pennsylvania State University, State College, PA 16801, USA}

\author{Marcus Hultmark}
\address{Princeton University, Princeton, NJ 08544, USA}

%\maketitle

\begin{abstract}
The wake of a horizontal-axis wind turbine was studied at $Re_D=4\times10^6$ with the aim of revealing the effects of the tip speed ratio, $\lambda$, on the wake. Tip speed ratios of $4<\lambda<7$ were investigated and measurements were acquired up to 6.5 diameters downstream of the turbine. Through an investigation of the turbulent statistics, it is shown that the wake recovery was accelerated due to the higher turbulence levels associated with lower tip speed ratios. The energy spectra indicate that larger broadband turbulence levels at lower tip speed ratios contributes to a more rapidly recovering wake. Wake meandering and a coherent core structure were also studied, and it is shown that these flow features are tip speed ratio invariant, when considering their Strouhal numbers. This finding contradicts some previous studies regarding the core structure, indicating that the structure was formed by a bulk rotor geometric feature, rather than by the rotating blades. Finally, the core structure was shown to persist farther into the near wake with decreasing tip speed ratio. The structure's lifetime is hypothesized to be related to its strength relative to the turbulence in the core, which decreases with increasing tip speed ratio.
\end{abstract}

\begin{keyword}
wind turbine \sep high Reynolds number \sep hot-wire \sep tip vortex \sep wake meandering \sep wake recovery
\end{keyword}

%{\bf MSC Codes }  {\it(Optional)} Please enter your MSC Codes here
\end{frontmatter}

%\linenumbers
\section{Introduction}
\label{sec:intro}
When wind turbines are organized as wind farms, it is common for some turbines to be in the wakes of others. A momentum deficit will be present in the inflow of those turbines as energy has been extracted by the upstream turbines, and consequently, the downstream turbine will be unable to extract as much energy from the wind as the upstream one. Therefore, the total power output and energy density of a wind farm depends on these turbine-turbine interactions.

Due to the complex geometry and large size of modern wind turbines, highly turbulent flow is expected in the wake, in combination with coherent flow structures generated by the rotor. This combination adds to our challenges in predicting and modeling these flows and systems. The flow structures play an important role in the wake recovery, which in turn limits the power density that can be achieved in a wind farm. At the tip of the blades, tip vortices are formed and shed into the wake in a helical fashion as they travel downstream, due to the blades' rotation \cite{porte2020}. Flow structures confined to the core of the wake are also commonly present, and sometimes consist of discrete root vortices shed from the roots of the blades \citep{sherry2013interaction,foti2016wake} or a hub vortex which is commonly characterized by a precessing helical vortex \citep{iungo2013linear,kang2014onset}. In addition, in some experiments a low frequency behavior has been observed in the wake, known as wake meandering. The mechanisms responsible for wake meandering are still being debated, but some theories include large-scale convection in the Atmospheric Boundary Layer (ABL), \citep{larsen2007dynamic,espana2011spatial} or bluff body-like shedding \citep{medici2006measurements}. The interactions between all three of these structures (tip vortex, wake meandering, and coherent core feature) will dictate how the wake will recover its momentum \citep{okulov2007stability,kang2014onset,foti2016wake}. Therefore, a strong understanding of how these structures are affected by different flow conditions, such as Reynolds number, tip speed ratio, turbulence intensity, etc., is imperative for more energy-efficient wind farms. There are two fundamental time scales governing the flow features in the wake: the turbine's angular velocity, $\omega$, and the convective timescale $D/U_\infty$, where, $D$ is a length scale associated with the geometry of the wake generator, here we will use the turbine diameter, and $U_\infty$ is the freestream velocity. A common nondimensional parameter characterizing the effect of the turbine's rotation rate is the tip speed ratio, $\lambda=\omega D\left({2U_\infty}\right)^{-1}$. Tip speed ratio is one of the most important non-dimensional parameters governing wind turbine aerodynamics and the wake features. The tip speed ratio can be though of as the ratio of the speed of the tip to the speed of the incoming flow, or as the ratio of the convective to rotational time scales.
%\citep{troldborg2010numerical,whale2000experimental,medici2006measurements}.

In addition, $\lambda$ has a strong effect on the turbine's performance metrics, such as the thrust $C_T=T{\left(0.5\rho A{U_{\infty}}^2\right)}^{-1}$, and power, $C_P=P{\left(0.5\rho A{U_{\infty}}^3\right)}^{-1}$, coefficients. Here, $T$ is the thrust force generated by the turbine, $\rho$ is the fluid density, $A$ is the area swept by the turbine, and $P$ is the power generated by the turbine. These performance metrics have a significant effect on the wake, as they characterize the power and momentum extraction from the flow. For example, $C_T$ has been shown to have an effect on wake evolution \citep{whale2000experimental} and shape \citep{eggleston1987wind}. The power coefficient is also expected to influence the wake, but its influence is primarily due to its dependence on $\lambda$. The tip speed ratio sets the local angle of attack  along the blade's span, and in turn, this affects the forces on the blades. The integral sum of the forces at each local airfoil section is responsible for the generation of power on a turbine. In turn, the local angle of attack of the airfoil section will set the vorticity shed into the wake, hence leading to a bulk wake effect.
%$C_P$ is also expected to influence the wake through its relationship with $\lambda$. Turbines have an optimal tip speed ratio, $\lambda_o$, which is the tip speed ratio for which the power coefficient, $C_P$, is at its maximum. For $\lambda<\lambda_o$, the angle of attack, $\alpha$, of the blades closest to the core is large. This large $\alpha$ leads to stall and large drag along the blades in the core, resulting in a more pronounced velocity deficit in the core. For $\lambda=\lambda_o$, the energy extraction along the blade span is optimized, such that a large velocity deficit will be found in the near wake. For $\lambda>\lambda_o$, $\alpha$ is small, and in some cases, negative, especially near the center of the turbine. This small or negative $\lambda$ feeds momentum into the core of the wake.

Many previous studies have investigated the effects of $\lambda$ on the wakes of wind turbines, but only a few of them have done so at Reynolds numbers, $Re_D=\rho U_\infty D \mu^{-1}$, relevant to modern wind turbines. Here, $\mu$ is the dynamic viscosity. As discussed above, local airfoil behavior, such as lift, drag, and stall, along a turbine's blade has an effect on the wake. Reynolds number has been shown to have an effect on an airfoil's lift curve, specifically in delaying stall for increasing Reynolds numbers \cite{devinant2002,alam2010,llorente2014,pires2016,brunner2021study}. In addition, it is well known that the Reynolds number affects transition \citep{tani1969,narasimha1985laminar}, laminar separation bubbles \citep{gaster1967structure}, and boundary layer statistics \citep{samie2018fully}. Changes in separation bubbles and boundary layer behavior will manifest as Reynolds number dependent stall behavior on a turbine's blade, leading to a Reynolds number dependent near wake. Therefore, in order to properly understand the effects of $\lambda$ on a turbine's wake, it is also important to maintain high Reynolds numbers. Although previous studies have found a lack of Reynolds number effects on the wake's mean velocity statistics \citep{chamorro2012reynolds,pique2022laboratory}, there is a lack of well resolved measurements conducted at high Reynolds numbers and at varying tip speed ratios. In the following section, the effects of rotation on the mean statistics, tip vortex, wake meandering, and coherent core structures are summarized.

\subsection{Effects of rotation}
The wake statistics, specifically the mean velocity deficit and the variance, are significantly influenced by the tip speed ratio, due to its strong effect on $C_T$. In general, increasing $C_T$, increases the velocity deficit in the near wake, which has been observed in previous studies \citep{sherry2013characterisation,yang2019wake,el2022numerical}. In addition, it has been hypothesized that with increasing $\lambda$, the turbine begins to increasingly resemble a solid disk, further explaining a larger deficit with increasing $\lambda$ \citep{sherry2013characterisation}. However, other studies have observed a larger velocity deficit for smaller $\lambda$ and $C_T$ up to four diameters downstream \citep{bastankhah2016experimental}. An explanation for this contradiction is the smaller angle of attack experienced by the blades near the core at higher $\lambda$, which leads to a speed-up region in the core \citep{bastankhah2015wind}. It is important to note that the detailed relationship between $\lambda$ and $C_T$ is unique to every turbine design. In addition to the mean velocity, the turbulence statistics are of utmost importance, both for the evolution of the wake itself, and due to the additional loads they generate on downstream turbines. As with the mean velocity deficit, the axial velocity variance is expected to have a dependence on $C_T$ because of stronger velocity gradients associated with higher velocity deficits \citep{ceccotti2016effect}.

The tip speed ratio has also been shown to affect the previously mentioned flow structures. The tip vortices have been shown to experience a more rapid collapse with increasing tip speed ratio \citep{troldborg2010numerical,sherry2013interaction,lignarolo2014experimental,sorensen2015simulation,sarlak2016assessment,el2022numerical}. The likely explanation for this behavior is the decrease in the tip vortex pitch, or the distance between the successive vortex cores of the helical vortex system, with increasing $\lambda$ \citep{wood1992wake,ebert1999near,ebert2001near,sherry2013interaction}. Helical vortex instabilities were studied extensively by \cite{widnall1972stability}. A critical finding relevant to wind turbines was that with a small enough pitch, mutual inductance of the helical vortex is the most likely unstable mode. Therefore, this Biot-Savart driven mutual inductance would occur more closely to the turbine plane with decreasing pitch/increasing $\lambda$ \citep{felli2011mechanisms}. 

Furthermore, the influence of tip speed ratio on wake meandering has also been studied. Wake meandering is commonly identified as a convection dominated low frequency event, that has been thought to require multiple diameters to develop. Thus, meandering is commonly characterized by a Strouhal number, $St=fD{U_{\infty}}^{-1}$, where $f$ is the frequency of the meandering. Values range from 0.12 \citep{medici2006measurements} to values on the order of 0.3 \citep{chamorro2013interaction,howard2015statistics}. The Strouhal number has been shown to be independent of $\lambda$ in the past \citep{yang2019wake}, but a large enough $\lambda$ is required to reach this invariant regime \citep{medici2006measurements}.

Much like the effect on the tip vortices, the pitch of the root or hub vortices will decrease with increasing $\lambda$ \citep{felli2011mechanisms}. Therefore, a more rapid collapse of the root vortex is also expected at higher $\lambda$, due to the same mechanisms responsible for the collapse of the tip vortex. Also, the root vortices have been found to become unstable before the tip vortices \citep{sherry2013characterisation}, which could be due to interactions with the nacelle boundary layer and consequent shear layer \citep{sherry2013interaction}. Instead of discrete root vortices, numerous studies have identified a precessing helical hub vortex in the wake core \citep{iungo2013linear,viola2014prediction,kang2014onset,ashton2016hub}. The precessing core feature has been shown to have a low frequency timescale associated with it, but unlike wake meandering, there is still some debate over its dependence on the turbine’s rotation rate. In \cite{iungo2013linear}, the ratio of the frequency of the hub vortex, $f_{hub}$, versus the turbine's angular velocity, $f_{rot}$, is relatively constant with tip speed ratio, and a numerical study that validated those findings found $f_{hub}/f_{rot}=0.32$ \citep{viola2014prediction}. However, in another study, the timescale of the hub vortex was found to not have a direct relationship with the turbine’s angular velocity, e.g. $f_{hub}/f_{rot}$ is not a constant \citep{ashton2016hub}. Therefore, questions still remain regarding the influence of the turbine's rotation on the hub vortex instability, a structure that is expected to be heavily dependent on the turbine’s detailed geometry \citep{kang2014onset}.

To summarize, a wind turbine's intrinsic rotation has an effect on its wake's bulk features, such as velocity deficit, and geometry-dependent flow structures, such as the tip vortex. Despite previous efforts to characterize the effects of rotation on a wind turbine's wake, there remains a lack of high Reynolds number studies on the topic. One objective of this investigation is to provide a tip speed ratio parametric study conducted at Reynolds numbers relevant to modern wind turbines. In this study, wake measurements of a model turbine are acquired for a range of downstream distances, $0.77<x/D<6.52$, and a range of tip speed ratios, $4<\lambda<7$. The experiments were conducted at a constant Reynolds number, $Re_D=4\times10^6$, which is at least an order of magnitude greater than previous wake experiments.

\section{Experimental Setup} \label{sec:methodology}

The experimental investigation presented herein was conducted in the High Reynolds number Test Facility (HRTF) at Princeton University. Rather than solely using the flow speed to increase Reynolds number, which is the case in a conventional wind tunnel, the HRTF allows the static pressure, and thus fluid density, to be increased to obtain a desired $Re_D$. Pressures up to 238 bar and freestream velocities up to 10 m/s are realizable in this flow facility. All tests were acquired at $Re_D=4\times10^6$, but with different combinations of $U_\infty$ and $\rho$. Dynamic similarity dictates that the non-dimensionalized statistical moments and spectra collapse as long as the $Re_D$ and $\lambda$ remain constant, as has been shown previously by \cite{pique2022laboratory,pique2022dominant}. Upstream of the turbine, the flow is conditioned to produce a uniform inflow for the turbine. The free stream turbulence intensity was $0.7\%$. As the current study was performed using the same wind turbine and instrumentation setup as used by \cite{miller2019,pique2022laboratory,pique2022dominant}, only a summary of the setup is presented here. For a more detailed description see those references. 

\begin{figure}
  \centerline{\includegraphics[width=\linewidth,keepaspectratio]{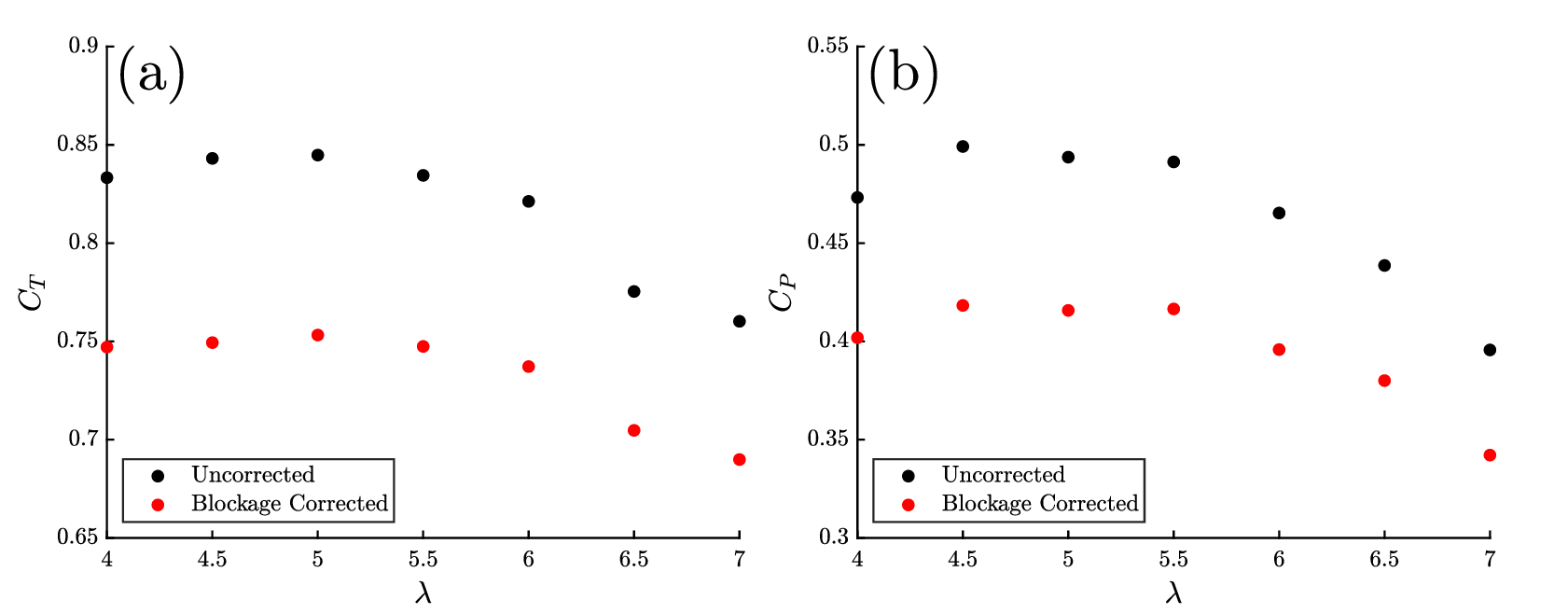}}
  \caption{$C_T$ (a) and $C_P$ (b) curves of the studied turbine at $Re_D=4\times10^6$. Uncorrected values are represented by black dots and blockage corrected values are represented by red dots. Blockage corrections were made following \cite{bahaj2007power}. Due to the qualitative nature of the analysis, $\lambda$ was not corrected for blockage.}
\label{fig:CtCpCurve}
\end{figure}

Thrust measurements were acquired using a JR3 Inc. six-axis load cell (model 75E20A4, 200N range) that was located at the bottom of the turbine tower, outside of the test section but inside the pressurized environment. All thrust data was acquired at 1000~Hz. An in-line torque transducer (Magtrol Inc., TM-305) was used to acquire both torque and angular velocity data. All data acquired from the torque transducer was sampled at 200kHz. Due to the self-starting nature of the turbine, a magnetic hysteresis brake (Magtrol Inc. AHB-3)  was used to control the rotational speed of the turbine. Freestream velocity data was acquired using a Pitot-static tube that was located 0.74m upstream of the turbine. Static pressure was measured using a pressure transducer (Omega Engineering Inc., PX419)  and temperature was measured using a resistance temperature detector. $C_T$ and $C_P$ curves can be found in Figure \ref{fig:CtCpCurve} over the range of $\lambda$ tested. Uncertainty in $C_T$ never exceeded 6.8\%, $C_P$ never exceeded 9.3\%, and $\lambda$ never exceeded 2.7\%. 

Velocity measurements in the wake were acquired using hot-wire anemometry. Due to the high $Re_D$ and reduced physical size of the wake, compared to a field-scale turbine, conventional hot-wire anemometry would be insufficient for measuring the smallest turbulent length scales. To resolve a greater range of length scales in the turbine's wake, a nano-scale thermal anemometry probe (NSTAP) was used \citep{vallikivi2014fabrication,fan2015nanoscale}. The NSTAP is fabricated at the Princeton Micro and Nano Fabrication Center (MNFC) using standard micro-electromechanical system (MEMS) techniques. The wire is composed of platinum and has dimensions of 60\textmu m $\times$ 2\textmu m $\times$ 100nm. A silicon structure supports the freestanding wire. With the use of a commercial operating circuit (Dantec Dynamics A/S, Streamline), the NSTAP was operated in constant-temperature mode (CTA). Calibrations of the NSTAP were performed before and after every measurement of a velocity profile. Temperature changes never exceeded 0.4$^{\circ}$C over the course of a wake measurement, so any temperature effects were deemed negligible. The repeatability in the hot-wire measurements was less than 0.15\%, estimated from the maximum difference between the pre- and post-calibration datasets.

\begin{figure}
  \centerline{\includegraphics[width=\linewidth,keepaspectratio]{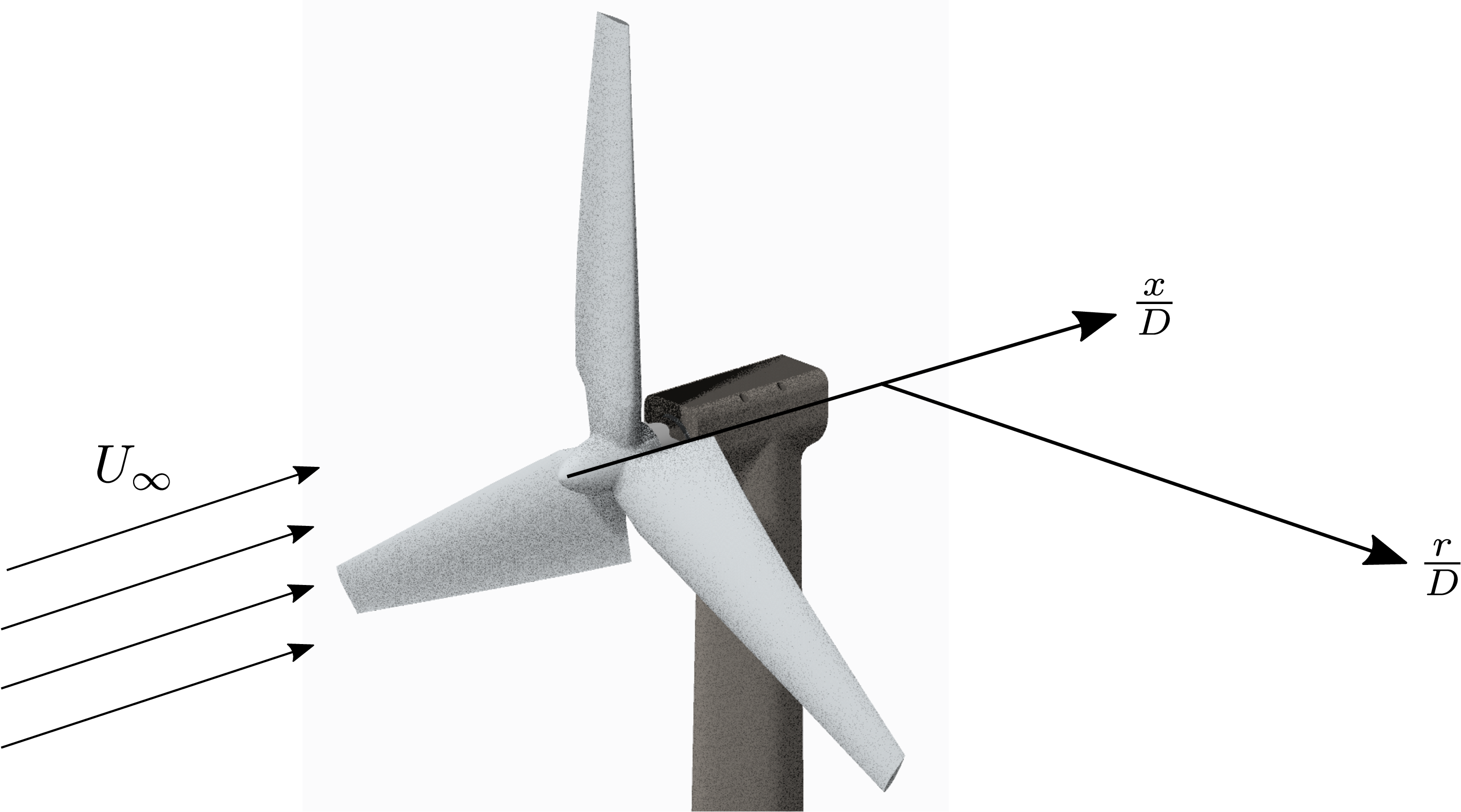}}
  \caption{Coordinate system of the presented wake results. Measurements were acquired along the $r/D$ axis at different streamwise positions along the $x/D$ axis. The turbine rotates clockwise.}
\label{fig:turbineCoord}
\end{figure}

Wake measurements were conducted along the spanwise plane, $-0.86<r/D<0.86$, for at least 200 rotations of the turbine at each spanwise position. The coordinate system used in this study is presented in Figure \ref{fig:turbineCoord}. Here, $x/D$ denotes the axial direction, and $r/D$ denotes the spanwise direction. 81 points are sampled at $x/D$=0.77 and 1.02 in order to improve the spanwise resolution of the tip vortex. 39 points were sampled at all other downstream positions. Mean axial velocity deficit and variance profiles are presented to aid a discussion on wake recovery and the effect of dominant flow structures on the wake's evolution. Spectral data are also analyzed and presented to further the discussion on the effects of turbine geometry and rotation on the wake evolution.
\section{Results}\label{sec:results}
As discussed briefly in Section \ref{sec:intro}, $C_T$ is expected to have an effect on the wake. As a result, it can be difficult to differentiate the sources of wake behavior due to the tip speed ratio or the thrust coefficient due to the intrinsic $\lambda-C_T$ relationship found for every turbine, as shown in Figure \ref{fig:CtCpCurve}a. However, for the current turbine, the maximum change in $C_T$ for $4<\lambda<6$ is no more than 2.9\%. Therefore, any trends observed in this range must primarily be an effect of the tip speed ratio, and thus the rotational effects. When considering the entire $\lambda$ range tested $\left(4<\lambda<7\right)$, the difference in $C_T$ is no more than 11.1\%. Therefore, trends that appear only over this range of $\lambda$, can be difficult to determine the root cause of. However, if a trend is observed for $4<\lambda<6$ and continues for $\lambda>6$, the authors still consider this a tip speed ratio effect.

The results and discussion presented here are a natural extension to the previously conducted high Reynolds number parametric studies of the same model turbine \cite{pique2022laboratory,pique2022dominant}. The tip speed ratio is one of the few operating conditions that can be toggled to control and predict turbine performance in a wind farm. Therefore, a tip speed ratio parametric study conducted at high Reynolds number is necessary to further improve wind farm operation and power output predictions. Some of the key findings of the previous studies were that the wake was populated by tip vortices that dominated the near wake ($x/D<2.02$), wake meandering ($St=0.3$) was identified at all downstream positions up to 5.5 diameters, and a low-frequency wake core structure ($St=0.6$) was observed only at the most upstream location $x/D=0.77$. Furthermore, the tip vortex was shown to have a broadband effect on the turbulent energy content which resulted in a reduced inertial subrange. The goal of the current study was to investigate how the wake is affected by varying tip speed ratio, at high Reynolds numbers, and whether it leads to changes in the evolution of the aforementioned wake structures or of the wake as a whole.
\subsection{Characterizing the wake} \label{sec:wakeStats}
\subsubsection{Wake recovery trends with tip speed ratio} \label{sec:defProfiles}
\begin{figure*}
  \centerline{\includegraphics[width=\linewidth,keepaspectratio]{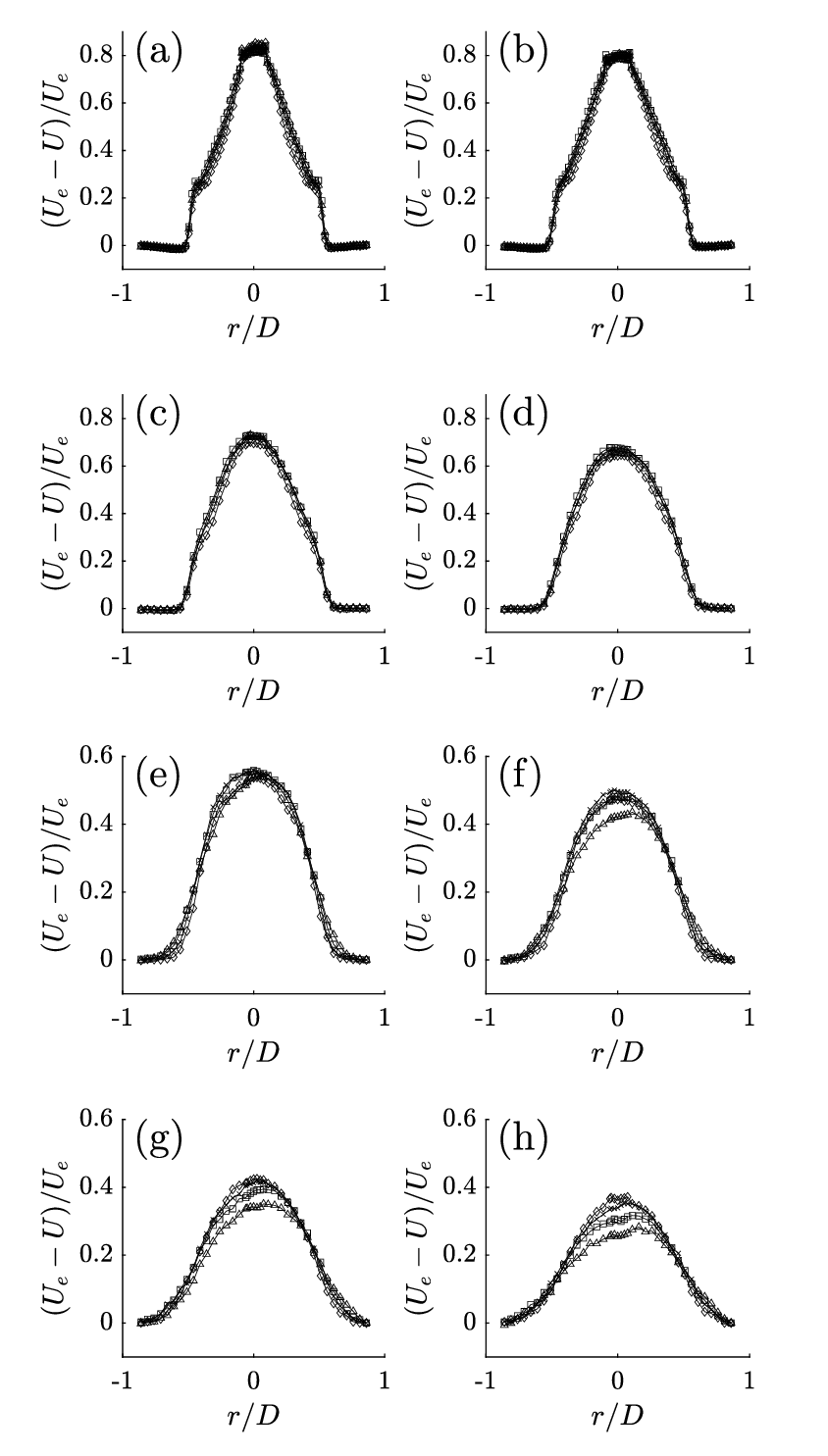}}
  \caption{Axial velocity deficit profiles across the tested downstream distances of $x/D=$0.77(a), 1.02(b), 1.52(c), 2.02(d), 3.52(e), 4.52(f), 5.52(g), 6.52(h) across all $\lambda$. $\lambda=4$($\bigtriangleup$), $\lambda=5$($\Box$), $\lambda=6$($\times$), $\lambda=7$($\Diamond$)}
\label{fig:deficitCompiled}
\end{figure*}

\begin{figure*}
  \centerline{\includegraphics[width=\linewidth,keepaspectratio]{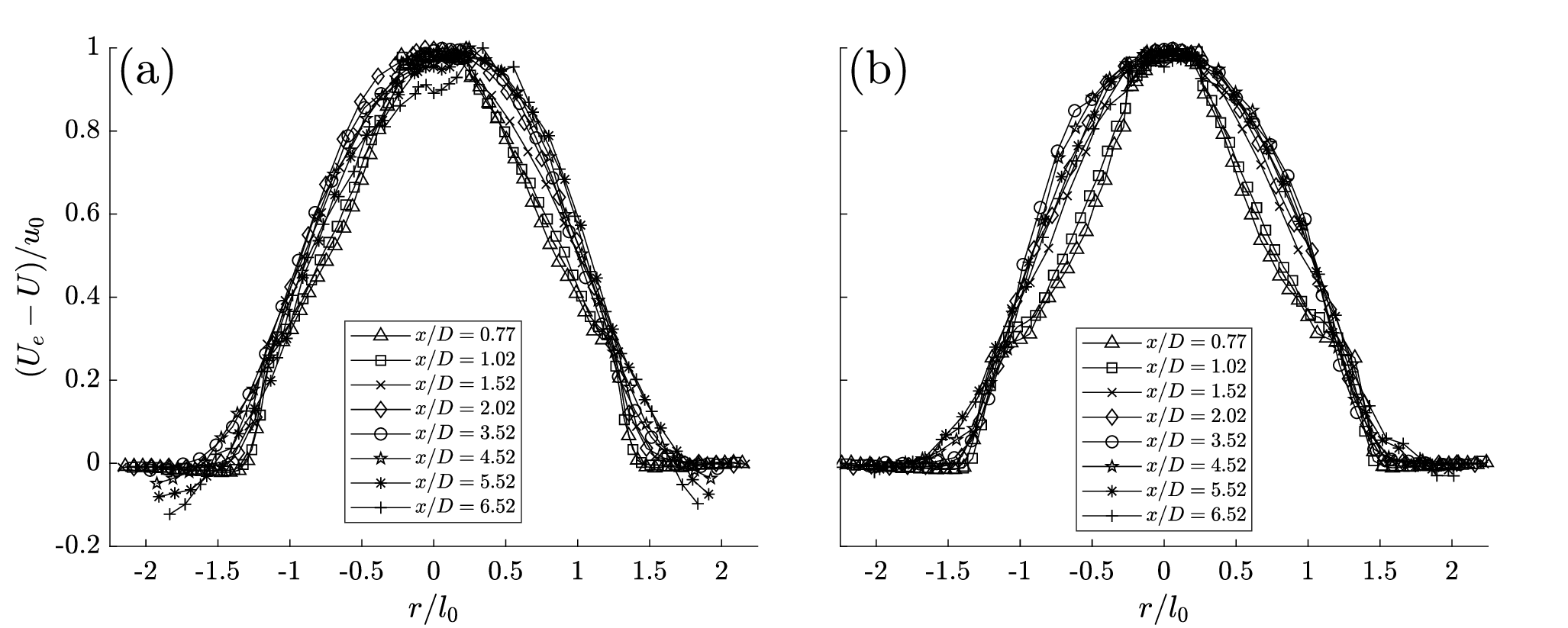}}
  \caption{Axial velocity deficit profiles across $0.77<x/D<6.52$ for $\lambda=4$ in (a) and $\lambda=7$ in (b). Profiles are nondimensionalized by classical self similar scales,  the deficit velocity ($u_0$) and the half-width ($l_0$).}
\label{fig:uDef_xD}
\end{figure*}

Profiles of the mean deficit velocity are shown in Figure \ref{fig:deficitCompiled} for the entire range of downstream locations and $\lambda$ tested. To account for blockage effects, the deficit velocity is calculated as the difference between local velocity and the velocity outside of the wake at the tested downstream position, $U_e$. First, there appears to be a change in the general shape of the deficit profiles in the range of $1.02<x/D<1.52$. To visualize this change, the profiles when nondimensionalized by conventional self-similar length scales, the half-width ($l_0$) and the deficit velocity ($u_0=U_e-U(r=0)$), as discussed by \cite{townsend}, are shown in Figure \ref{fig:uDef_xD}. The half-width is the spanwise distance between the centerline ($r/D=0$) and the point at which the velocity deficit is $\frac{u_0}{2}$. It is clear that the wake transitions out of the very near wake somewhere in the range $1.02<x/D<1.52$, with a self-similar profile appearing beyond that. Figure \ref{fig:deficitCompiled} shows that the mean deficit velocity is independent of $\lambda$ for $0.77\leq x/D\leq3.52$. However, for $x/D>3.52$, the collapse starts to deteriorate with the $\lambda=4$ and $5$ cases showing a more rapid deviation, and the lowest $\lambda$ case showing signs of a faster rate of wake recovery. The $\lambda=6$ and $7$ cases continue to collapse well with each other, but a greater downstream range may be needed to evaluate if this continues to hold further downstream. 

The mean velocity deficit profile can be used to determine the thrust coefficient, given that the profile is acquired far enough downstream of the rotor, where the radial velocity and pressures can be assumed negligible. Given the consistent inflow conditions in the current study, the wakes of the studied turbine would be expected to collapse with enough downstream distance, due to the relatively small changes in $C_T$. However, for the presented turbine, the deficit profiles lose their self-similar behavior for $x/D>3.52$ and fail to obtain them by $x/D=6.52$. Thus, the point at which one can calculate $C_T$ from the deficit profiles, without additional information, is beyond the tested range.

\subsubsection{Turbulence effects due to tip speed ratio} \label{sec:varianceProf}
Profiles of the variance of the axial velocity are shown below in Figure \ref{fig:TKECompiled}. At the most upstream locations, $x/D\leq2.02$, the profiles exhibit four distinct peaks, whereas further downstream they exhibit only two peaks. The outer peaks, in the upstream locations, are indicators of the tip vortices and the inner peaks are signatures of an axisymmetric shear layer in the wake core. There is an asymmetry in the inner peak magnitudes, which are still unexplained but a commonly observed phenomena also in other wake studies \citep{odemark2013stability,schumann2013experimental,vinnes2022far}. Interestingly, for the two most upstream locations ($x/D=0.77$ and $1.02$), the entire profiles collapse with $\lambda$. Further downstream (starting at $x/D=1.52$) the collapse starts to deteriorate and the higher $\lambda$ cases see a decrease in the magnitude of the outer peaks (tip vortex signatures), whereas the inner region remains largely unchanged. The four-peak profile can still be observed up to $x/D=2.02$, but with decreasing magnitude of the the outer peaks corresponding to the decreasing strength of the tip vortices with increasing downstream distance \citep{hu2012dynamic}. Increasing $\lambda$ tightens the pitch of the helical vortex and reduces the strength of the tip vortices due to smaller angles of attack along the blade span. Both of these effects are expected to yield faster decay of the tip vortices with decreasing outer peaks, and lower variances in general, as a result.  

\begin{figure*}
\centerline{\includegraphics[width=\linewidth,keepaspectratio]{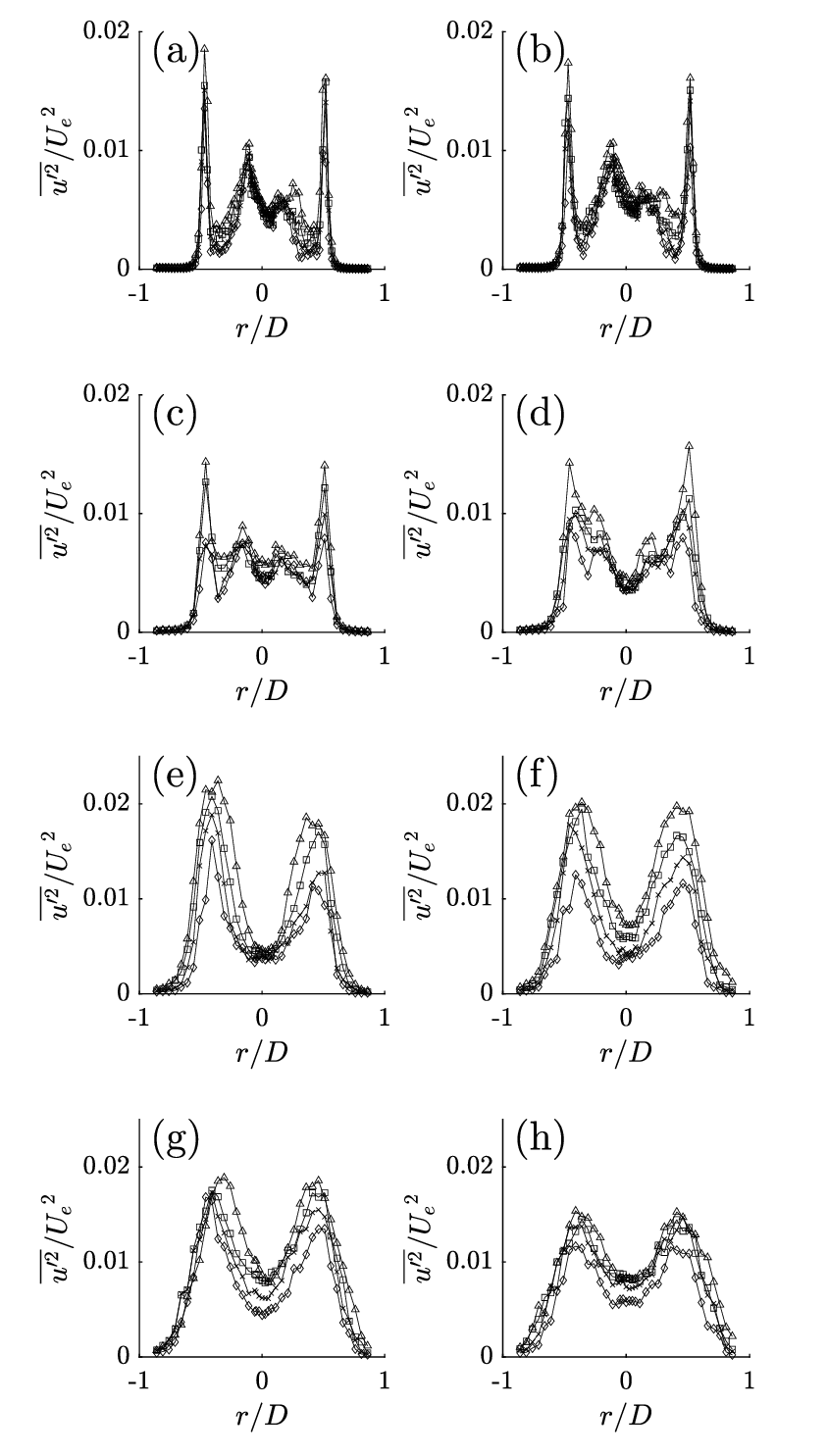}}
\caption{Axial velocity variance profiles across the tested downstream distances of $x/D=$0.77(a), 1.02(b), 1.52(c), 2.02(d), 3.52(e), 4.52(f), 5.52(g), 6.52(h) across all $\lambda$. $\lambda=4$($\bigtriangleup$), $\lambda=5$($\Box$), $\lambda=6$($\times$), $\lambda=7$($\Diamond$)}
\label{fig:TKECompiled}
\end{figure*}

When a tip vortex collapses, a burst of turbulent kinetic energy has been observed \citep{el2022numerical} and is thought to be due to increased entrainment from the `leapfrogging' of the tip vortices \citep{lignarolo2015tip}. At $x/D=3.52$ (shown in Figure \ref{fig:TKECompiled}e) the profiles have transitioned to a two-peak profile and have a greater magnitude than the peaks at $x/D=2.02$, supporting previous findings of an increase in turbulent energy after the collapse of the tip vortex. The transition to a two-peak profile has been observed in past studies \citep{odemark2013stability,tedds2014near,sarlak2016assessment}, and indicates the point at which the near wake structures have collapsed to form a single annular shear layer \citep{foti2016wake}. The near wake is typically defined as the flow closest to the wake generator (turbine in this case) that is still dominated by turbine-dependent structures, such as tip and root vortices. The intermediate wake is a region where most of the near wake structures have disappeared, but self-similar statistics are still not obtainable \citep{pique2022laboratory}, and the far wake is where self-similarity is expected. As shown in \cite{johansson2006far}, the two-peak profile of an axisymmetric, non-rotating wake can persist into the far wake region. Therefore, based on the variance profiles, the near wake can be expected to extend to at least $x/D=2.02$ with the intermediate wake starting at a maximum of $x/D=3.52$. 

The failure of collapse that was first observed in the near wake variance profiles, starting at $x/D=1.52$, can also be observed in the intermediate wake, showing that the effect of $\lambda$ and the breakdown of the tip vortex extends into the intermediate wake. In other words, initial conditions have an effect not only on the near wake, but also on the intermediate wake. For all downstream positions in the intermediate wake up to 5.52 diameters downstream, increasing $\lambda$ correlates with a decrease in the magnitude of the two dominant peaks. As shown in Figure \ref{fig:TKECompiled}, the greater the tip speed ratio, the smaller the variance, with the most notable trend for $1.52<x/D<5.52$. The initial burst of turbulent kinetic energy after the collapse of the tip vortex will decay due to recovery mechanisms, such as entrainment from surrounding regions. Therefore, it is likely that the higher turbulence levels in the vicinity of the tip vortices, will accelerate wake recovery for lower $\lambda$, as shown in Figure \ref{fig:deficitCompiled}. Based solely on the variance profiles, it is difficult to determine whether the tip vortex alone is responsible for the $\lambda$ dependence observed. However, the dependence on $\lambda$ decreases further downstream, to the point at $x/D=6.52$, the variance profiles have almost collapsed once again.

As discussed in Section \ref{sec:results}, $C_T$ remains relatively unchanged over the range of tested $\lambda$. For any one $C_T$ value, there is not a single unique force distribution along the blade. The force distribution along a rotor's blade is dependent on the angle of attack of the local airfoil section, which in turn depends on the turbine's tip speed ratio. In addition, the local angle of attack will determine the vorticity shed into the wake, thus affecting the strength and distribution of vortical structures being shed into the wake. Therefore, it is expected that the turbine's rotation will affect both the local vorticity and the dominant wake flow structures (such as tip or hub vortices). For the studied turbine, vortical structures formed in the near wake show significant rotational effects throughout the tested regime, specifically the turbulence levels in the near and intermediate wake associated with the tip vortex as shown in Figure \ref{fig:TKECompiled}. These effects cannot be attributed to $C_T$, as it is almost constant. A thorough investigation of the energy spectra will provide more clues as to the $\lambda$ relationship and its effects on wake evolution.

\subsection{Periodic structures}
As discussed in Section \ref{sec:varianceProf}, the turbulence generated by the tip vortices is hypothesized to be an important mechanism for wake recovery. However, it is difficult to determine the effects of other dominant flow structures, such as wake meandering and coherent core phenomena, on the wake recovery and their relationship with the tip speed ratio. Therefore, a spectral analysis was conducted to determine how $\lambda$ affects the different wake features. Here, the spectra are non-dimensionalized by $f_{rot}$ or by the convection time scale (yielding a Strouhal number, $St$). The different non-dimensionalizations are used to determine if a periodic structure is governed by the rotation of the blades or by flow convection. 
\subsubsection{Tip vortex}
\begin{figure*}
  \centerline{\includegraphics[width=\linewidth,keepaspectratio]{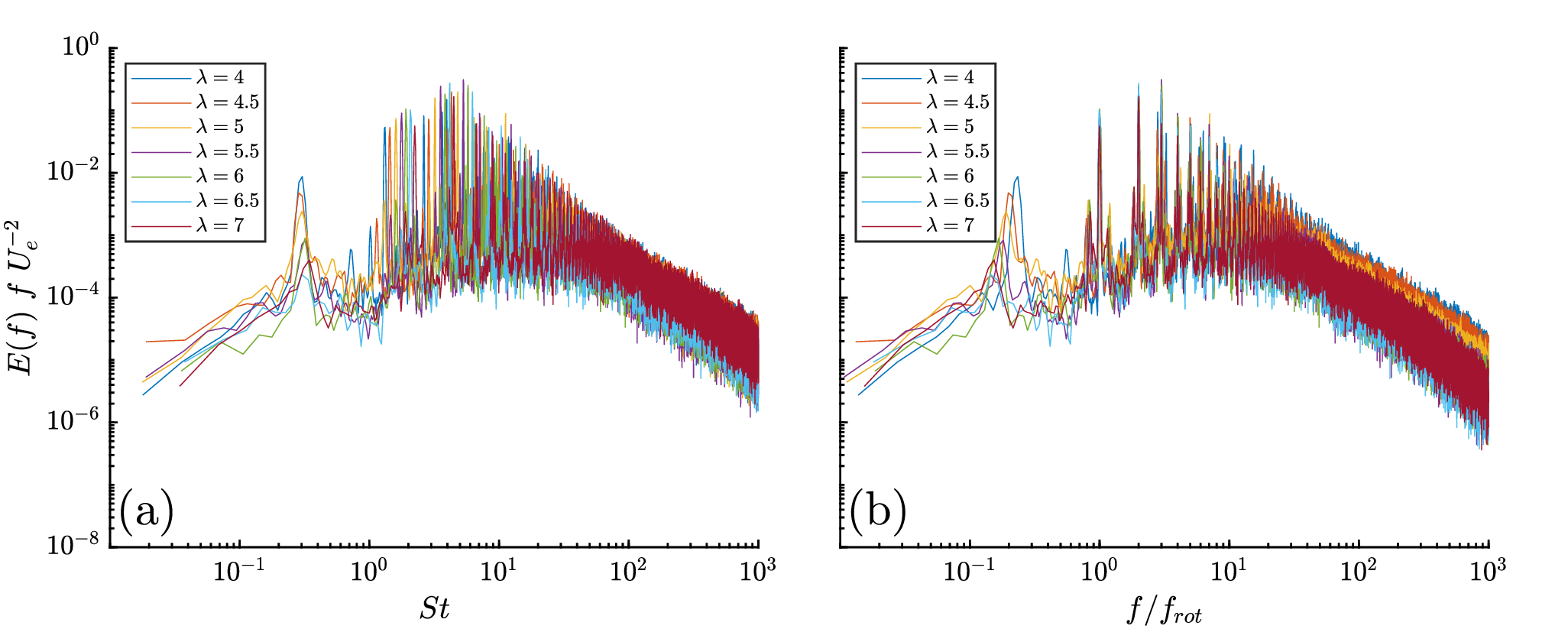}}
  \caption{Premultiplied spectrum at $x/D=0.77$ and $r/D=0.519$, a location near the tip vortex. The frequency is nondimensionalized by a Strouhal number (a) or by the rotational frequency of the turbine, $f_{rot}$ (b).}
\label{fig:spectrumTipVortex_TSRCompare}
\end{figure*}

In Figure \ref{fig:spectrumTipVortex_TSRCompare}, spectra near the tip vortex location ($r/D=0.519$ and $x/D=0.77$) are shown. In Figure \ref{fig:spectrumTipVortex_TSRCompare}a the frequencies are non-dimensionalized as a Strouhal number, with a peak present at $St\approx0.3$, which corresponds to the wake meandering reported in previous studies. The $St$ of that peak remains close to constant across all $\lambda$, which agrees with the earlier observations of wake meandering \citep{medici2006measurements,yang2019wake}. Also, the magnitude of the wake meandering peak decreases with increasing $\lambda$. In Figure \ref{fig:spectrumTipVortex_TSRCompare}b, the frequencies are instead non-dimensionalized by $f_{rot}$. Here, the influence of the tip vortex can be more clearly observed. Strong peaks are identified at multiples of $f_{rot}$ ($f_{rot}$, $2f_{rot}$, and $3f_{rot}$). It can also be seen that the tip vortices have a strong broadband effect for frequencies greater than $f_{rot}$. This range is similar across all $\lambda$ tested for $x/D=0.77$. The turbine's tip vortices push the inertial subrange to higher reduced frequencies, with an increased broadband energy content. Therefore, the broadband energy strengthening effect of the tip vortex further supports the previous hypothesis that the tip vortex has a lasting effect on wake recovery. 

\begin{figure*}
  \centerline{\includegraphics[width=\linewidth,keepaspectratio]{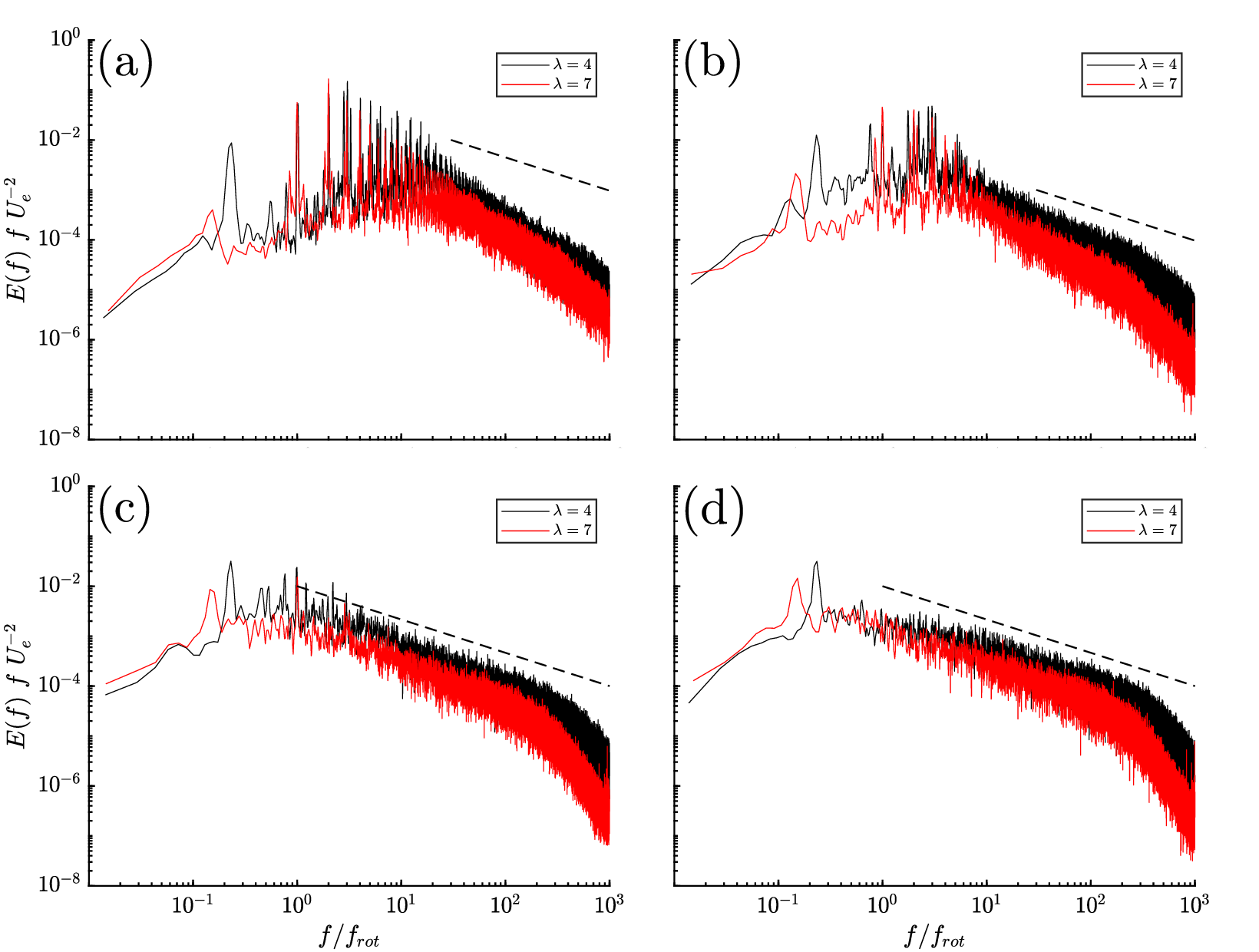}}
  \caption{Premultiplied spectra at (a) $x/D=0.77$, (b) 2.02, (c) 3.52, and (d) 6.52 . The spectra are obtained near the location of the tip vortex, but due to the different spanwise resolutions for the downstream positions, the radial position is slightly different. In (a), $r/D=0.519$ and in (b-d), $r/D=0.51$. Dashed lines represent a -2/3 slope, a reference to the inertial subrange in premultiplied spectrum scaling.}
\label{fig:tipVortexStrength}
\end{figure*}

From Section \ref{sec:varianceProf}, the magnitude of the tip vortex signature in the variance was found to decrease with increasing $\lambda$. Since there was a monotonic trend with $\lambda$, only the spectra at $\lambda=4$ and $7$ are shown in Figure \ref{fig:tipVortexStrength} to more clearly illustrate the trends with tip speed ratio. When considering the near wake, it is clear that the tip speed's ratio influence over all reduced frequencies is much stronger at $x/D=2.02$ than at $x/D=0.77$. As shown in Figure \ref{fig:tipVortexStrength}, the overall energy content near the blade tip decreases with increasing tip speed ratio. This trend of decreasing energy with increasing $\lambda$ corresponds to the lower variance magnitudes observed in Figure \ref{sec:varianceProf}. At the most upstream location, $x/D=0.77$, there is only a weak $\lambda$ dependence (and possibly none at the lower reduced frequencies), agreeing well with the collapse of the axial variance profiles for the same downstream position. The broadband effect of the tip vortices alters the turbulence and shifts the inertial subrange (as identified by the dashed lines in Figure \ref{fig:tipVortexStrength}) to smaller lengthscales, but the affected frequency range decreases with increasing downstream distance, within the near wake.

Premultiplied spectra at $x/D=3.52$ and $x/D=6.52$ are shown in Figure \ref{fig:tipVortexStrength}c and d, respectively. The lower $\lambda$ case shows a greater energy content across all scales at $x/D=3.52$, supporting the hypothesis that the turbulence generated by the tip vortices in the near wake also affects the intermediate wake. This trend was discussed in Section \ref{sec:varianceProf} and is found in the greater magnitude variance profiles for smaller $\lambda$ of Figure \ref{fig:TKECompiled}e. At $x/D=6.52$, shown in Figure \ref{fig:tipVortexStrength}d, the trend with $\lambda$ is less pronounced especially for $f/f_{rot}<3$, the region most affected by the tip vortex, representing a large fraction of the total energy (recall the collapse of the variance profiles at the $x/D=6.52$ in Figure \ref{fig:TKECompiled}h). 
% suggesting that at this point, the flow is becoming independent of the initial conditions, such as the $\lambda$-dependent turbulence. This observation is supported by the closer collapse of the variance profiles at the same downstream position as shown in Figure \ref{fig:TKECompiled}h.
% \begin{figure*}
%   \centerline{\includegraphics[width=\linewidth,keepaspectratio]{Figures/TipVortexStrength_IntermediateWake_xD.eps}}
%   \caption{Premultiplied spectrum at $x/D=3.52$ (a) and $x/D=6.52$ (b) at a location near the tip vortex ($r/D=0.51$).}
% \label{fig:tipVortexStrength_intermediateWake}
% \end{figure*}

\subsubsection{Persistence of coherent wake core structures} \label{sec:wakeCore}
\begin{figure*}
  \centerline{\includegraphics[width=\linewidth,keepaspectratio]{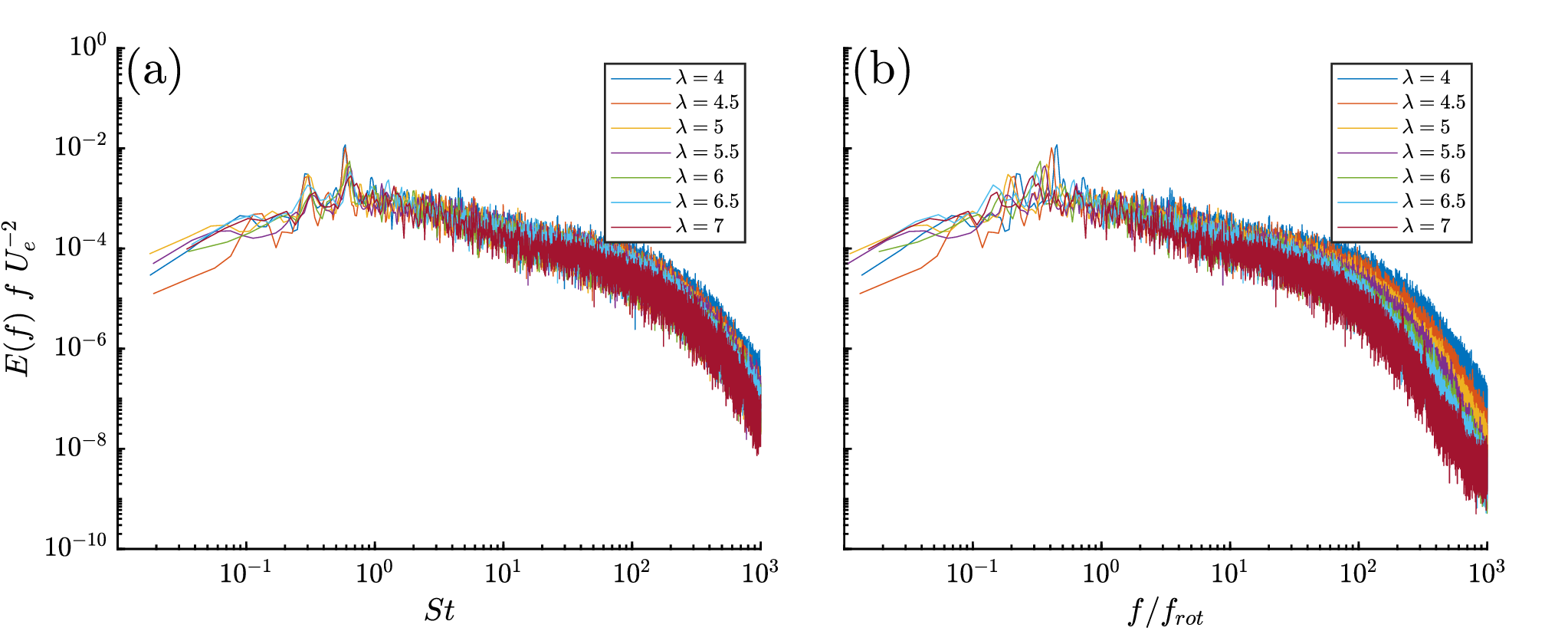}}
  \caption{Premultiplied spectrum at $x/D=0.77$ and $r/D=0.007$, a location in the wake core. The frequency is nondimensionalized by a Strouhal number (a) or by the rotational frequency of the turbine, $f_{rot}$ (b)}
\label{fig:spectrumWakeCore_TSRCompare}
\end{figure*}

In Figure \ref{fig:spectrumWakeCore_TSRCompare}, the premultiplied spectra in the wake core, $r/D=0.007$ and $x/D=0.77$ are shown for different $\lambda$. Two clear peaks can be observed, one at $St=0.3$ and one at $St=0.6$. Wake meandering is typically associated with a $St=0.3$, and as mentioned in Section \ref{sec:results}, the $St=0.6$ peak is a signature of shedding in the wake core. As discussed in Section \ref{sec:intro}, wake core structures have been identified and are hypothesized to be signatures of the root vortex or a precessing helical vortex. From Figure \ref{fig:spectrumWakeCore_TSRCompare}b, there is no collapse associated with any of the harmonics of $f_{rot}$, and therefore it is unlikely that there are any root vortices at $x/D=0.77$ for any tip speed ratios tested. A precessing helical vortex in the wake core has been identified in previous studies, which showed that there is a dependence on $f_{rot}$ \citep{iungo2013linear}. However, for the case of the presented turbine, no such structure is observed. The $St=0.6$ signature remains constant for all $\lambda$ tested, suggesting that a bulk feature of the turbine, rather than blade-driven formation mechanisms, is driving the wake core structure observed at $St=0.6$.

\begin{figure*}
  \centerline{\includegraphics[width=\linewidth,keepaspectratio]{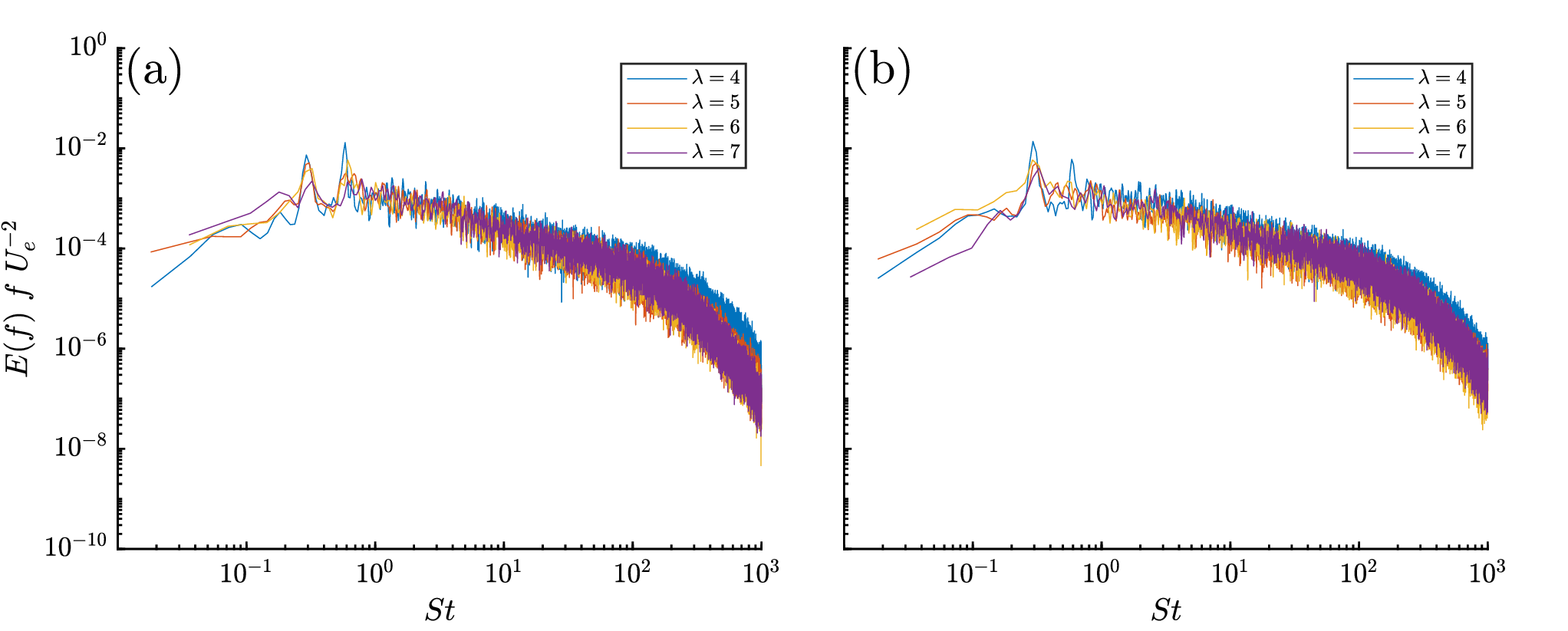}}
  \caption{Premultiplied spectrum at (a) $x/D=1.02$ and $r/D=0.007$ and (b) $x/D=1.52$ and $r/D=0.01$. There is a slight mismatch in the spanwise position due to a different spanwise resolution between the two downstream locations.}
\label{fig:spectrumWakeCore_nearWake}
\end{figure*}

As discussed in \cite{pique2022dominant}, the structure at $St=0.6$ is shortlived, only present in the very near wake of the turbine. Figure \ref{fig:spectrumWakeCore_nearWake} shows the premultiplied spectra at $x/D=1.02$ and $x/D=1.52$. Once again, the wake meandering and wake core structures have a constant $St$ with changing $\lambda$. However, it is clear that with increasing $\lambda$ the magnitude of the $St=0.6$ event decreases, until the point where the signature disappears. In Figure \ref{fig:spectrumWakeCore_nearWake}b, at $x/D=1.52$, the $St=0.6$ signature is only evident at the lowest $\lambda$. For all further downstream locations, the signature of the $St=0.6$ event has disappeared. Therefore, the lifetime of the core structure is dependent on the tip speed ratio, where a lower $\lambda$ prolongs the downstream extent of the structure. Also, unlike the tip vortex, the tip speed ratio does not have a strong effect on the energy content of any of the other scales in the core. This observation was expected due to the strong degree of collapse of the variance profiles in the wake core in the near wake as discussed in Section \ref{sec:varianceProf}.

\subsubsection{Spanwise extent of the wake core signature}
\begin{figure*}
  \centerline{\includegraphics[width=\linewidth,keepaspectratio]{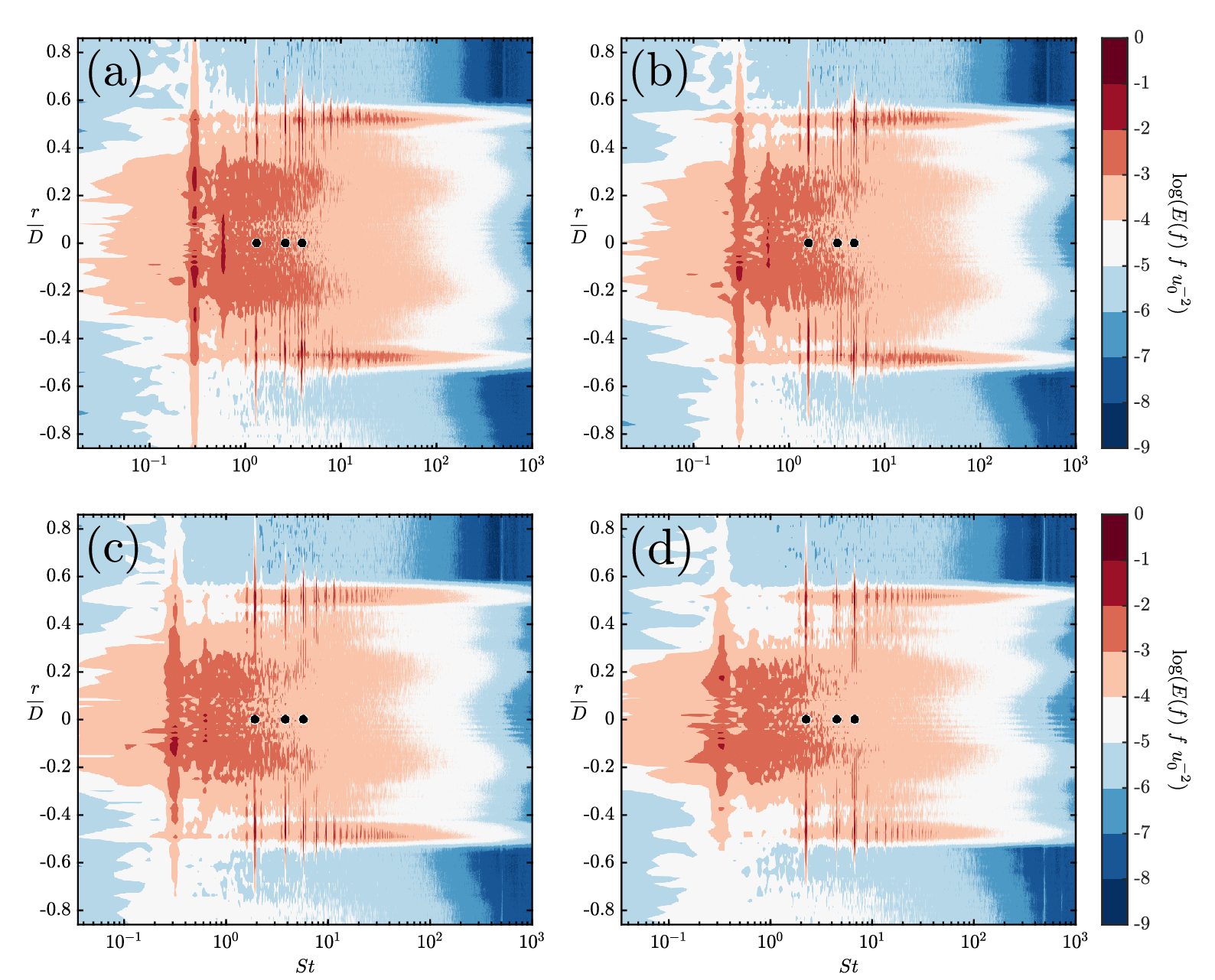}}
  \caption{Premultiplied spectrum displaying spanwise extent of dominant flow features at $x/D=0.77$ for $\lambda=4$(a), 5(b), 6(c), and 7(d). Black dots correspond to $St$ for $f_{rot}, 2f_{rot}, 3f_{rot}$.}
\label{fig:contourSpectrum}
\end{figure*}

\begin{figure*}
  \centerline{\includegraphics[width=\linewidth,keepaspectratio]{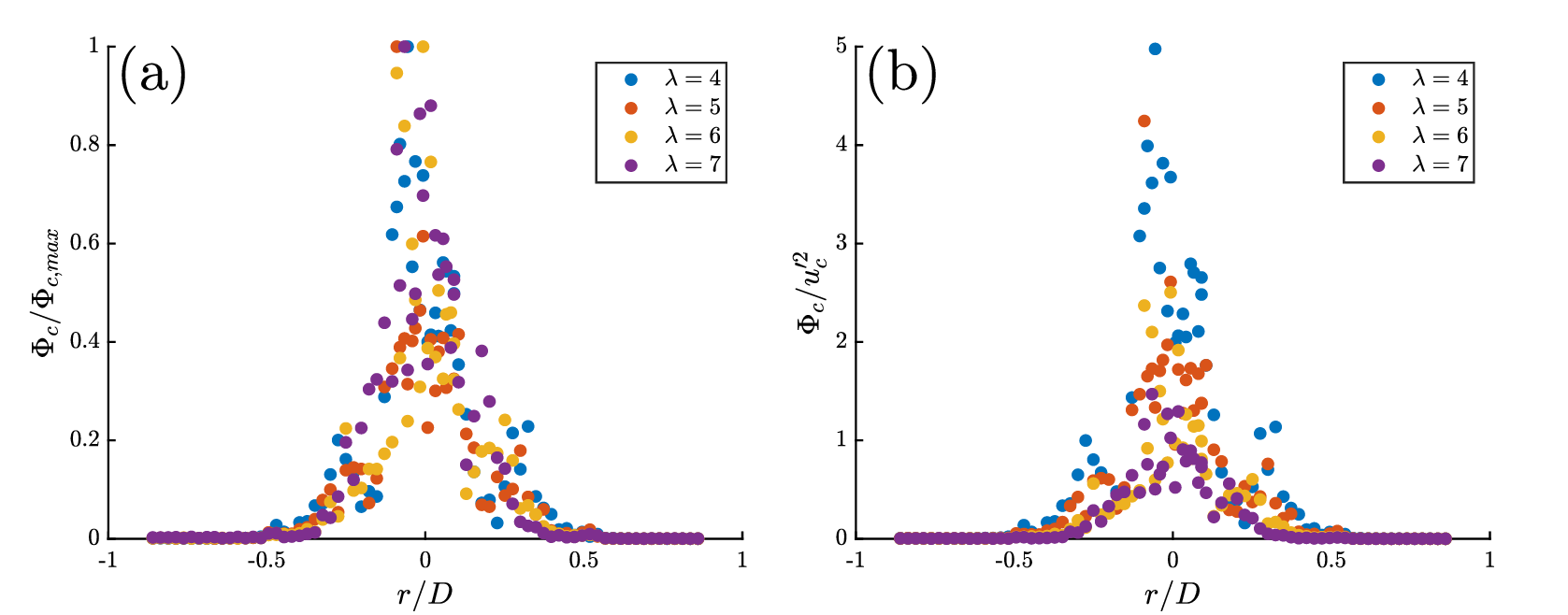}}
  \caption{Peaks of the premultiplied spectrum at $x/D=0.77$ associated with the maximums in the range of $0.55<St<0.65$, $\Phi_c$. In (a), the premultiplied spectrum nondimensionalized by the greatest signature magnitude associated with the core shedding, $\Phi_{c,max}$. In (b), the premultiplied spectrum are nondimensionalized by the variance at the centerline, ${u_c'^{2}}$.}
\label{fig:coreStructuresSignatures}
\end{figure*}

In this study, the $St=0.6$ signature has been shown to be affected by the tip speed ratio, such that it has a smaller downstream extent and is weaker with increasing $\lambda$. To further visualize the trend between $\lambda$ and the strength of the low frequency features (wake meandering and the core structure), the premultiplied spectrum presented as a contour plot are shown in Figure \ref{fig:contourSpectrum}. A principal feature of the spectrum is that increasing tip speed ratio leads to a reduction in the strength and spanwise influence of the wake meandering and core structure features. To better visualize the trend between $\lambda$ and the core structure's strength, the peaks in the premultiplied spectrum in the Strouhal number range of $0.55<St<0.65$, $\Phi_c$, have been identified in Figure \ref{fig:coreStructuresSignatures}. In Figure \ref{fig:coreStructuresSignatures}a the premultiplied spectrum is non-dimensionalized by the maximum value of the spectrum, $\Phi_{c,max}$, to better visualize the spanwise extent of the wake core signature. In Figure \ref{fig:coreStructuresSignatures}b the premultiplied spectrum is non-dimensionalized by the variance at the centerline, ${u_c'^{2}}$, to better visualize the strength of the core structure relative to turbulence in the core. From Figure \ref{fig:coreStructuresSignatures}a, the wake core signature is found to have a similar radial extent, independent of $\lambda$. Despite the tip speed ratio invariance of the spanwise lengthscale of the wake core signature, the tip speed ratio has a strong effect on the strength of this structure. In Figure \ref{fig:coreStructuresSignatures}b the strength of the wake core structure is shown to decrease with increasing tip speed ratio, and can be expected to have a smaller effect on the flow in the core. This trend helps support the previous finding from Section \ref{sec:wakeCore} that the downstream lifetime of the wake core signature decreases with increasing $\lambda$: for the studied turbine, the strength of the wake core signature relative to the turbulence in the core decreases with increasing $\lambda$, leading to a faster collapse.

%As has been previously discussed, an increasing $\lambda$ is expected to lead to an increase in the turbines relative solidity \cite{}. This argument was expanded in \cite{pique2022dominant} in which the  mechanism responsible core shedding of the same turbine was due to the high relative solidity of the turbine's core. Therefore, it was expected that the spanwise extent of the core shedding behavior would increase with increasing $\lambda$ due to the increase in the relative solidity. 

%At the surface level, this observation contradicts the findings of \cite{pique2022dominant} that the wake core structure is due to the turbine's solidity because one might expect that with increasing $\lambda$, the turbine would act more like a solid disk, leading to an increase in the spanwise extent at which the solidity is large enough to initiate bluff body-like shedding. However, as discussed in Section \ref{sec:results}, the $C_T$ experiences little change in the range of $4<\lambda<6$, indicating that the relative solidity of the turbine is no different across the range. Therefore, the inavariance of the wake core structure's spanwise lengthscale  with $\lambda$ is supported by the $C_T$ invariance for $4<\lambda<6$.

\section{Conclusions}
In this study, the wakes of a wind turbine were investigated at $Re_D=4\times10^6$ for a range of tip speed ratios, $4<\lambda<7$. The near and intermediate wake regions were studied using nanoscale hot-wire anemometry. It was shown that tip speed ratio has a significant effect on the wake dynamics, which could not be explained by the small changes in the $C_T$ for changing $\lambda$. It was further shown that mean deficit velocity profiles were invariant in the near wake, whereas the profiles in the intermediate wake showed a dependence on $\lambda$, with a smaller $\lambda$ leading to a more rapid wake recovery. This observation contradicted some previous findings, but an investigation of the axial variance profiles indicated stronger turbulence levels for lower tip speed ratios across most of the downstream positions tested, specifically $1.52<x/D<5.52$. However, the variance profiles alone were insufficient for deciphering which flow features contribute to the modified turbulence levels.

Within the near wake, the turbulent energy as measured by the variance profiles, was found to decrease with increasing $\lambda$ in the proximity of the tip vortex, whereas it was close to independent of $\lambda$ at the wake core for the vast majority of scales. Through spectral analysis, the energy content of the turbulent length scales in the proximity of the tip vortex were investigated to determine which scales contributed to the higher turbulence levels at smaller tip speed ratios. The tip vortex was shown to have a broadband effect across a wide range of scales, which elevates the turbulence levels in its proximity. However, at the farthest downstream point of the near wake ($x/D=2.02$), the energy content of the tip vortex scales was found to be greater for smaller $\lambda$, coinciding with the same trends found in the variance. The trends in the energy content of the tip vortex scales persisted into the intermediate wake, again supporting the same trends found in the variance profiles for the same region. The relationship between the tip vortex scale energy content and $\lambda$ supports the hypothesis that a smaller tip speed ratio accelerates wake recovery within the intermediate wake due to a stronger tip vortex and its more energetic turbulence. The wake recovery trend with tip vortex strength indicates the importance of accurately resolving the tip vortex and its broadband effect in future modeling efforts.

%Through spectral analysis, the wake meandering and core features were first investigated. The wake core showed a spectral peak at $St=0.6$, independently of $\lambda$, which contradicts previous findings and is a strong indication that the core structure for the studied turbine is not strictly governed by the turbine blades, but is rather due to a bulk geometric feature of the unique turbine. This finding is important for the wind community as the wake core structure is of great interest, and one that we don't fully understand. In addition, within the near wake, the turbulent energy as measured by the variance profiles, was found to decrease with increasing $\lambda$ in the proximity of the tip vortex, whereas it was close to independent of $\lambda$ at the wake core. With respect to the tip vortex, the $\lambda$ effects observed in the spectrum were affecting all turbulent length scales in the near wake ($1.52<x/D<2.02$), whereas in the intermediate wake, only the smaller length scales were affected, with the larger, energy containing length scales, were independent of $\lambda$. Therefore, it is hypothesized that the stronger turbulence levels generated over a wide range of lengthscales in the proximity of the tip vortex with decreasing $\lambda$ is a contributing factor to the more rapid wake recovery associated with lower tip speed ratio. 

Wake meandering and the wake core structure were found to dominate the low-frequency content of the wake. The wake core structure was found to have a spectral peak at $St=0.6$, independent of $\lambda$, contradicting previous findings. The $\lambda$ invariance is a strong indication that the core structure is not strictly governed by the turbine blades, but is rather due to a bulk geometric feature of the unique turbine. The strength of wake meandering and the wake core signature are both found to be dependent on the tip speed ratio: their magnitudes decrease with increasing $\lambda$. Furthermore, the wake core structure is shown to have a greater downstream extent with decreasing tip speed ratio. Although the spanwise extent of the wake core structure was found to be invariant with $\lambda$, the relative strength of the core structure across the span was found to be dependent; decreasing $\lambda$ produced a stronger core structure. The stronger wake core structure relative to the turbulence in the core at lower tip speed ratio was determined to be the reason for its increased longevity.

\section{Declaration of Competing Interest}
The authors declare that they have no known competing financial interests or personal relationships that could have appeared to influence the work reported in this paper.

\section{Data Availability}
Data are available from the corresponding author, AP, upon reasonable request.

\section{Author Contributions}
\textbf{Alexander Piqu\'e}: methodology, data curation, writing-original draft preparation, formal analysis, investigation, visualization \textbf{Mark A. Miller}: writing-review and editing, methodology \textbf{Marcus Hultmark}: writing-review and editing, supervision 

\section{Funding Sources}
This work was supported by the National Science Foundation under Grant No. CBET 1652583.

\section{Acknowledgements}
The authors would like to acknowledge the support of the National Science Foundation under Grant No. CBET 1652583 (Program Manager Ron Joslin).

\bibliographystyle{elsarticle-num}
\bibliography{References.bib}% common bib file

\end{document}